\documentclass[preprint,12pt]{elsarticle}

\usepackage{color}
\usepackage{amssymb}
\usepackage{amsmath}

\journal{Materials Research Bulletin}

\begin{document}

\begin{frontmatter}


\title{Micrometer thick single crystal iron-garnet films on a diamagnetic buffer layer for cryogenic applications}


\author[label1]{Kuzmichev~A.N\corref{cor1}} \ead{a.kuzmichev@rqc.ru}
\cortext[cor1]{corresponding author}
\author[label1,label2]{Vetoshko~P.M.} 
\author[label2]{Pavluk~E.I.} 
\author[label2]{Holin~A.A.} 
\author[label1,label3]{Knyazev~G.A.} 
\author[label1,label3]{Kaminskiy~A.S.} 
\author[label1]{Demirchan~S.S.} 
\author[label4]{Tyumenev R.} 
\author[label4]{Kalashnikov D.S.} 
\author[label4]{Stolyarov V.S.} 
\author[label1,label3]{Belotelov~V.I.} 

\affiliation[label1]{organization={Russian Quantum Center},
            city={Moscow},
            postcode={121205}, 
            country={Russian Federation}}

\affiliation[label2]{organization={V.I. Vernadsky Crimean Federal University},
            city={Simferopol},
            postcode={295007}, 
            country={Russian Federation}}
            
\affiliation[label3]{organization={Lomonosov Moscow State University, Faculty of Physics},
            city={Moscow},
            postcode={119991}, 
            country={Russian Federation}} 
            
\affiliation[label4]{organization={Moscow Institute of Physics and Technology},
            city={Moscow},
            postcode={141701}, 
            country={Russian Federation}}  
          
 Abstract
 \begin{abstract}
 This work advances the frontier of low-damping magnetic materials, directly addressing the demand for ultra-low-loss components in quantum computing and cryogenic electronics. Here we demonstrate a new approach to get single crystal micrometer-thick yttrium iron garnet (YIG) films with low damping through isolating and mitigating interfacial paramagnetic contributions of a paramagnetic substrate by a buffer-layer. The YIG films with the diamagnetic yttrium scandium gallium garnet buffer layer grown by liquid phase epitaxy on a gadolinium gallium substrate demonstrate homogeneity unprecedented for the thin planar YIG structures, yielding ferromagnetic resonance linewidths of 4.9 MHz at 4 K and 5.9 MHz at 16 mK, the lowest values reported to date. These results underscore the critical role of interfacial engineering in overcoming intrinsic material limitations, opening avenues for further optimization in spin-based technologies. 
\end{abstract}

\begin{graphicalabstract}
\includegraphics[width=10cm]{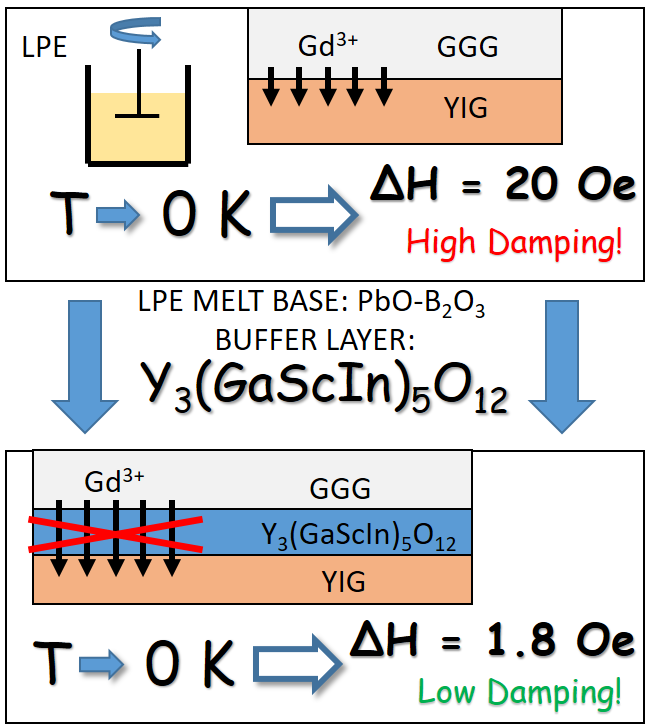}
\end{graphicalabstract}


\begin{highlights}
\item Record-low FMR linewidths of f \textbf{4.9 MHz at 4 K} and \textbf{5.9 MHz at 16 mK} -- the lowest
values reported to date. 
\item Novel ${\text{Y}_3\text{(GaScIn)}_5\text{O}_{12}}$ buffer layer engineered via LPE eliminates gadolinium diffusion, suppressing paramagnetic damping below 100~K.
\item Demonstrated quantum-ready performance at \textbf{16~mK} (dilution-refrigerator temperatures) for superconducting hybrid systems.
\item Lattice mismatch reduced to \textbf{0.0025~\AA} via ${\text{In}\textsuperscript{3+}}$ substitution, enabling near-spherical homogeneity in films.
\item Optimized LPE synthesis with ${PbO-B\textsubscript{2}O\textsubscript{3}}$ solution-melts enables industrial-scale production of high-purity films.
\end{highlights}


\end{frontmatter}



\section{Introduction}
\label{sec1}

Yttrium iron garnet (YIG) single crystals have long captivated researchers due to their exceptionally narrow ferromagnetic resonance (FMR) linewidth, a property pivotal to advancements in radio-frequency electronics. To date, YIG remains indispensable in radio-frequency devices such as filters, circulators, and phase shifters, owing to its superior insulating properties and low magnetic damping, which are critical for high-performance applications. A central question persists: how can magnetic losses in YIG be further minimized and what are the fundamental limits of this reduction? Early studies by Spencer, Kasuya, and others \cite{Spencer1959, Kasuya1961, Spencer1961} identified temperature-dependent damping mechanisms, with narrowing of the FMR linewidth at cryogenic temperatures. Subsequent work attributed these losses primarily to defects and impurities in bulk YIG crystals \cite{Seiden1964, Dionne1997, Dionne2000, Vittoria2010, Huebl2017}. However, the difficulty in handling and the high cost of YIG spheres spurred efforts to develop thin-film alternatives compatible with industrial-scale electronics \cite{Serga2010, Chumak2015}.

The rise of quantum magnonics and cryogenic spintronics has renewed urgency in optimizing YIG's damping properties at low temperatures, since long spin coherence times and compact device architectures are essential for quantum technologies and space-constrained cryogenic systems \cite{Nakamura2019, Nakamura2020, Novosad2020}. Epitaxial YIG films grown on gadolinium gallium garnet (GGG) substrates initially appeared promising, enabling long-range spin-wave propagation. However, interfacial challenges emerged, including lattice mismatch-induced strain, and, crucially, exchange and magnetic dipole interactions with the paramagnetic GGG substrate. GGG is a strong paramagnetic material with a significant stray field at cryogenic temperatures as shown in \cite{Serha2024}. Our prior research \cite{Kuzmichev2022} revealed an intermediate interfacial layer enriched with Gd and Fe ions, which dominates magnetic relaxation at temperatures below 100 K. The magnetic moment of this layer grows exponentially at low temperatures, introducing an additional damping that cannot be mitigated using conventional methods.

To address this, we proposed substituting GGG with diamagnetic substrates like \(\text{Y}_3\text{Sc}_2\text{Ga}_3\text{O}_{12}\) (YSGG). Although initial trials of growing YIG films on YSGG via liquid phase epitaxy (LPE) eliminated FMR frequency shifts below 50 K \cite{Kupchinskaya2024}, substrate imperfections hindered precise analysis of FMR linewidth behavior at cryogenic temperatures. Here, we present an alternative strategy: LPE growth of a gadolinium-free buffer layer on GGG prior to YIG growth. Although magnetron-sputtered buffer layers have reduced intrinsic damping by several times, inhomogeneous broadening limited FMR linewidths to 20 Oe at 5 K \cite{GUO2022}. In this work, our LPE-grown thick YIG films achieved exceptional homogeneity, yielding FMR linewidths of 1.82 Oe (3 GHz) at 4 K and 2.2 Oe (5.5 GHz) at 16 mK --- the lowest values reported to date.

\section{Sample preparation}
\label{sec2}

\begin{figure}[t]
\centering
\includegraphics[width=12cm]{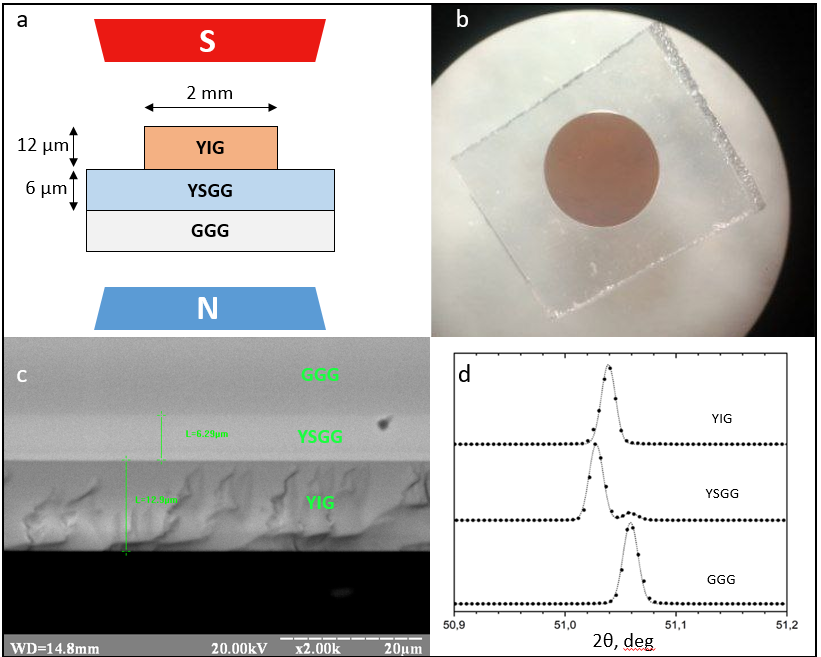}
\caption{(a) Structure of the sample with the buffer layer, (b) Photo of the sample with etched YIG element, (c) SEM image from scanning electron microscope, (d) Measured lattice mismatches with X-ray diffraction (XRD) in $\theta-2\theta$ geometry, focusing on the (444) Bragg reflections.}\label{fig1}
\end{figure}

To develop low-damping YIG films for cryogenic applications, we epitaxially grew and characterized a \(\text{Y}_3\text{(FeIn)}_5\text{O}_{12}\) (YIG) film incorporating a buffer-layer. The sample structure (Fig.~1a) consists of a 6~$\mu$m-thick \(\text{Y}_3\text{(GaScIn)}_5\text{O}_{12}\) buffer layer, grown via LPE on a GGG substrate with [111] crystallographic orientation, followed by a 12~$\mu$m-thick YIG layer deposited atop the buffer. Growth was performed in a three-zone furnace using a platinum crucible containing a PbO-based solution-melt, following established LPE protocols \cite{Shone1985}.  

As previously revealed \cite{Kuzmichev2022}, the diffusion of paramagnetic Gd\textsuperscript{3+} ions from the substrate into the adjacent part of the interface layer causes additional damping and ferromagnetic resonance (FMR) line broadening in iron-garnet films at cryogenic temperatures. Due to the lack of high-quality gadolinium-free yttrium-scandium gallium garnet (YSGG) substrates, a monocrystalline diamagnetic buffer layer was synthesized on a high-quality GGG substrate. The base composition for this layer was Y\textsubscript{3}(GaScIn)\textsubscript{5}O\textsubscript{12} garnet, adapted to match the lattice parameter of GGG. Commercial YSGG crystals with composition Y\textsubscript{3}Sc\textsubscript{2}Ga\textsubscript{3}O\textsubscript{12} typically have a lattice parameter \(a = 12.4569\) {\AA}, while GGG has a smaller parameter (\(a = 12.3729\) {\AA}). To minimize lattice mismatch, Sc\textsuperscript{3+} ions (\(R = 0.745\) {\AA}) in the octahedral sublattice were partially replaced by larger In\textsuperscript{3+} ions (\(R = 0.80\) {\AA}).

\begin{table}[ht]
\centering
\caption{Composition of the melt for synthesizing the Y\textsubscript{3}(GaScIn)\textsubscript{5}O\textsubscript{12} buffer layer.}
\label{tab:buffer}
\small
\setlength{\tabcolsep}{4pt}
\begin{tabular}{|l|rrrrrrr|}
\hline
Component & \multicolumn{1}{c}{PbO} & \multicolumn{1}{c}{B\textsubscript{2}O\textsubscript{3}} & \multicolumn{1}{c}{Y\textsubscript{2}O\textsubscript{3}} & \multicolumn{1}{c}{Sc\textsubscript{2}O\textsubscript{3}} & \multicolumn{1}{c}{Ga\textsubscript{2}O\textsubscript{3}} & \multicolumn{1}{c}{In\textsubscript{2}O\textsubscript{3}} & Total \\
\hline
Conc. (mol.\%) & 84.6 & 11.3 & 0.4 & 0.3 & 3.3 & 0.1 & 100.000 \\
Mass (g) & 2232.00 & 93.2 & 11.1 & 5.5 & 72.8 & 0.8 & 2315.3 \\
\hline
\end{tabular}
\end{table}

To reduce the lattice mismatch in Y\textsubscript{3}Fe\textsubscript{5}O\textsubscript{12} (YIG), In\textsuperscript{3+} ions (\(R = 0.80\) {\AA}) were introduced into the octahedral sublattice to partially substitute Fe\textsuperscript{3+}. The synthesis used a PbO--B\textsubscript{2}O\textsubscript{3} solution-melt. Adjustments in the lattice parameters and magnetic properties were achieved by controlling In\textsuperscript{3+} concentrations. The melt for Y\textsubscript{3}(InFe)\textsubscript{5}O\textsubscript{12} films contained PbO, B\textsubscript{2}O\textsubscript{3}, Fe\textsubscript{2}O\textsubscript{3}, Y\textsubscript{2}O\textsubscript{3}, and In\textsubscript{2}O\textsubscript{3} (Table 2). All oxides met high-purity standards (GOST 13867-68).

\begin{table}[ht]
\centering
\caption{Composition of the PbO-containing melt for synthesizing Y\textsubscript{3}(InFe)\textsubscript{5}O\textsubscript{12} films}
\label{tab:film}
\small
\setlength{\tabcolsep}{4pt}
\begin{tabular}{|l|cccccc|}
\hline
Component & PbO & B\textsubscript{2}O\textsubscript{3} & Fe\textsubscript{2}O\textsubscript{3} & Y\textsubscript{2}O\textsubscript{3} & In\textsubscript{2}O\textsubscript{3} & Total \\
\hline
Conc. (mol. \%) & 83.63 & 3.99 & 11.73 & 0.42 & 0.03 & 100.00 \\
Mass (g) & 2432.13 & 36.20 & 224.00 & 12.27 & 8.67 & 2713.27 \\
\hline
\end{tabular}
\end{table}

The YIG film was subsequently patterned into 2~mm-diameter discs using optical lithography and wet etching (Fig.~1b). A resist mask was first fabricated with a Heidelberg~mPG~101 maskless aligner, followed by selective etching in orthophosphoric acid to define the final sample shape.    

\section{Experiment setup}
\label{sec3}

\begin{figure}[t]
\centering
\includegraphics[width=12cm]{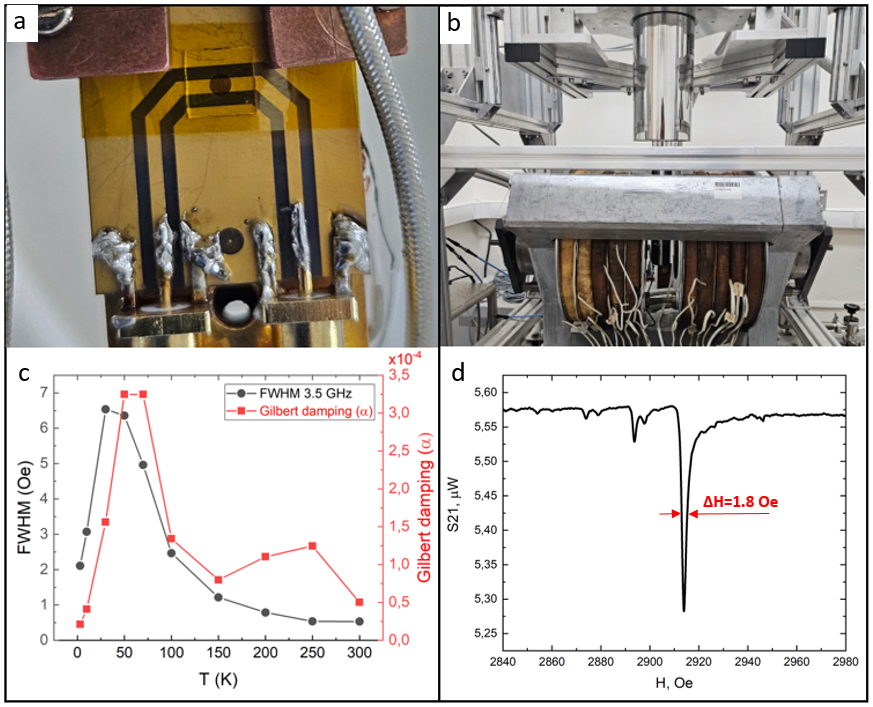}
\caption{(a) Coplanar waveguide line with YIG sample, (b) closed-loop cryocooler with electromagnet, (c) FWHM at 3.5 GHz and corresponding Gilbert damping constant dependence on temperature, (d) S21 FMR spectrum at 4 K with 3 GHz pumping frequency.}\label{fig2}
\end{figure}

Ferromagnetic resonance (FMR) measurements were performed at 300~K and 4~K using a coplanar waveguide structure comprising an \( \text{Al}_2\text{O}_3 \) dielectric substrate and a 1~$\mu$m-thick Cu metallization layer electroplated with Au (Fig.~2a). A closed-loop \( ^4\text{He} \) cryocooler coupled to an external electromagnet facilitated temperature control and magnetic field application during low-temperature experiments (Fig.~2b). Microwave frequency characterization was conducted using a Keysight~P5023A vector network analyzer (VNA).  

For ultralow-temperature measurements at 16~mK, a dilution refrigerator with superconducting magnet was used. The system utilized a hybrid \( ^3\text{He}\)-\( ^4\text{He} \) dilution refrigerator (Bluefors) integrated with a superconducting electromagnet (American Magnetics, Inc., AMI). Microwave signals were generated and detected using a Planar C1220 VNA.  

\section{Results and discussion}
\label{sec4}

\begin{figure}[t]
\centering
\includegraphics[width=12cm]{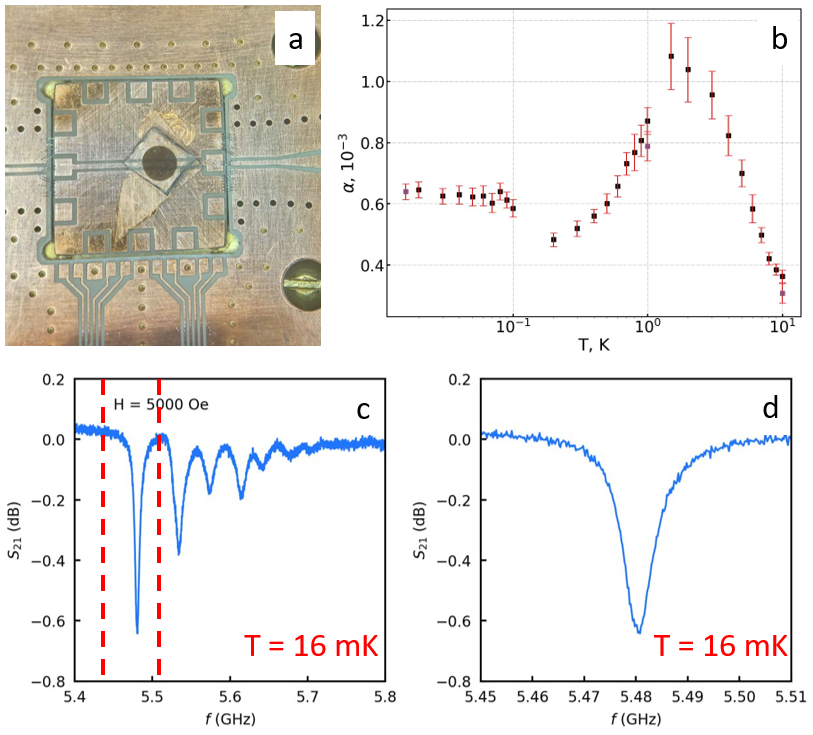}
\caption{(a) Photo of the YIG sample on the CPW insert for cryogenic measurements in a hybrid \( ^3\text{He}\)-\( ^4\text{He} \) dilution refrigerator, (b) Gilbert damping constant of the YIG sample at different temperatures, (c) S21 parameter spectrum representing FMR with 5000 Oe external magnetic field at 16 mK, (d) scaled part of S21 parameter spectrum marked with red dashed lines in (c).}\label{fig3}
\end{figure}

To assess the quality and uniformity of the \(\text{Y}_3\text{(ScGa)}_5\text{O}_{12}\) (YSGG) buffer layer and yttrium iron garnet (YIG) films grown on top, we performed cross-sectional analysis by cutting a portion of the sample post-growth and imaging it with a scanning electron microscope (SEM; REM-106I, Fig.~1c). Both the YSGG buffer and YIG film exhibited homogeneous thickness and interfacial integrity, confirming optimal growth conditions. 

A critical factor influencing the microwave performance of YIG films is the lattice mismatch parameter \(\Delta a = a_{\text{substrate}} - a_{\text{film}}\), where \(a\) denotes the lattice constant. Excessive \(\Delta a\) induces elastic strain, promoting magnetic anisotropy and inhomogeneous FMR line broadening. To quantify \(\Delta a\), we conducted X-ray diffraction (XRD) in \(\theta\text{-}2\theta\) geometry, focusing on the (444) Bragg reflections of the garnet structure (Fig.~1d). The measured lattice mismatches were \(\Delta a_{\text{GGG/YIG}}\) = 0.0045~\text{\AA} and \(\Delta a_{\text{YSGG/YIG}} = -0.0025~\text{\AA}\), with an experimental uncertainty of \(\pm 0.0005~\text{\AA}\). These values lie well below the threshold for significant strain-induced linewidth broadening, ensuring minimal impact on FMR line homogeneity.  

Fig.~2c displays the measured temperature dependence of the FMR linewidth along with the corresponding Gilbert damping coefficient. For each temperature point, FMR spectra were recorded at different frequencies. The damping coefficient was subsequently determined from the slope of the linear fit to the dependence of the resonance linewidth on the resonance frequency. The plots reveal a distinct line broadening and an increase in the damping coefficient around 50~K. This broadening is known to be associated with magnetic impurities and divalent iron ions \cite{Spencer1959, Kasuya1961, Spencer1961}. In particular, a clear decrease in the damping coefficient is observed in the temperature range from 250~K to 150~K, consistent with previously published studies on YIG spheres. 

Figure~2d presents a representative FMR spectrum recorded at 3~GHz and 4~K. The resonance exhibits a record-low linewidth of $\Delta H=1.8$~Oe for the films at 4~K, which is, to the best of our knowledge, the lowest value compared to prior literature. It corresponds to the FMR frequency linewidth of 4.9 MHz.

For the integration of YIG films into hybrid structures with superconducting qubits, low losses at sub-liquid-helium temperatures, specifically below 20~mK, are essential. This temperature regime is typical for superconducting qubit operation, as reducing the thermal noise energy (on the order of $k_{\text{B}}T$) is necessary to enhance state stability and readout fidelity. The sample was mounted within a dilution refrigerator stage, positioned inside a superconducting solenoid alongside a coplanar waveguide structure (Fig. 3a). To accommodate the coplanar waveguide insert, the sample substrate was cut on its sides. 

Fig.~3b shows the Gilbert damping constant in the temperature range of 16 mK to 10 K on a logarithmic scale. The behavior of the dependence is similar to that of FMR line width dependence in a YIG sphere measured by Nakamura \cite{Nakamura2014}. In this set up, the YIG sample diameter is significantly larger than the waveguide width. This geometry provokes additional modes at higher frequencies (Fig. 3c). Only the main mode was taken into account for the damping calculation. The FMR linewidth of the main mode at 16 mK with 5000 Oe external field is about 6 MHz (Fig. 3d), which corresponds to approximately 2.2 Oe at 5.5 GHz. The Q factor for the 16 mK resonance is 1.7 times larger than for the 4 K measurement ($\approx900$ vs $\approx600$, respectively).

A reduction in impurities concentration is also expected to decrease the FMR linewidth. Nevertheless, this demonstrates the potential of YIG epitaxial films for application in on-chip superconducting qubit circuits.

\section{Summary}
\label{sec5}

The application of micrometer-thick YIG films in cryogenic on-chip systems is promising for tuning, achieving non-reciprocity, filtering, periodic signal generation, and other functions. In this work we demonstrated that it is possible to have a FMR linewidth broadening as low as 6 MHz at cryogenic temperatures for the epitaxial micrometer-thick YIG films utilizing the buffer layer without paramagnetic ions. The results reported in this paper pave the way for obtaining high quality YIG films with cryogenic magnetic damping comparable to the record values achieved for the YIG spheres by LeCraw and others \cite{Spencer1959}. Key factors crucial for obtaining low damping YIG films are identified: high level of chemicals' purity, uniform and strain-free diamagnetic substrate. However, here we show that the widely spread paramagnetic GGG substrate can also be used if a diamagnetic buffer layer is epitaxially grown on top to avoid the harmful influence of the paramagnetic ions on FMR damping in YIG at low temperatures. 

\section*{Acknowledgments}
 This work was supported by the Russian Science Foundation (grant № 23-62-10024). Part of the cryogenic measurements was supported by the Ministry of Science and Higher Education of the Russian Federation (No. FSMG2023-0014).

\end{document}